\documentclass{aastex}
\usepackage{emulateapj5,natbib,psfig}

\newcommand{\oqkev}{$1\over4$~keV}
\newcommand{\tqkev}{$3\over4$~keV}
\newcommand{\rosat}{{\it ROSAT}}
\newcommand{\asca}{{\it ASCA}}
\newcommand{\rass}{{\it RASS}}
\def\lesssim{\mathrel{\hbox{\rlap{\hbox{\lower4pt\hbox{$\sim$}}}\hbox{$<$}}}}
\def\arcdeg{\hbox{$^\circ$}}
\def\arcmin{\hbox{$^\prime$}}

\shortauthors{Kuntz, Snowden, \& Mushotzky}
\shorttitle{WHIM-sical Observations}

\begin{document}

\title{X-ray Constraints on the Warm-Hot Intergalactic Medium}

\author{K. D. KUNTZ\altaffilmark{1},}
\affil {\normalsize {Department of Astronomy, University of Maryland, College Park, MD 20742}}
\author{S. L. SNOWDEN\altaffilmark{2,3} and R. F. MUSHOTZKY\altaffilmark{4}}
\affil {\normalsize {NASA Goddard Space Flight Center, Greenbelt, MD 20771}}

\altaffiltext{1}{E-mail-I:kuntz@astro.umd.edu}
\altaffiltext{2}{E-mail-I:snowden@lheavx.gsfc.nasa.gov}
\altaffiltext{3}{Universities Space Research Association}
\altaffiltext{4}{E-mail-I:richard@xray-5.gsfc.nasa.gov}

\begin{abstract}
Three observational constraints can be placed on 
a warm-hot intergalactic medium (WHIM)
using \rosat\ PSPC pointed and survey data,
the emission strength, the energy spectrum, and the fluctuation spectrum.
The upper limit to the emission strength of the WHIM is 
$7.5\pm1.0$ keV s$^{-1}$ cm$^{-2}$ sr$^{-1}$ keV$^{-1}$ in the \tqkev\ band,
an unknown portion of which value may be due to our own Galactic halo.
The spectral shape of the WHIM emission
can be described as thermal emission with $\log T=6.42$,
although the true spectrum is more likely to come from a range of temperatures.
The values of emission strength and spectral shape
are in reasonable agreement with hydrodynamical cosmological models.
The autocorrelation function in the 0.44 keV $< E <$1.21 keV band range, 
$w(\theta)$,
for the extragalactic soft X-ray background (SXRB)
which includes both the WHIM and contributions due to point sources,
is $\lesssim0.002$ for $10\arcmin < \theta < 20\arcmin$ in the \tqkev\ band.
This value is lower than the \citet{cmdhkfw2000} cosmological
model by a factor of $\sim5$,
but is still not inconsistent with cosmological models.
It is also found that the normalization of the extragalactic power law
component of the soft X-ray background spectrum
must be $9.5\pm0.9$ keV s$^{-1}$ cm$^{-2}$ sr$^{-1}$ keV$^{-1}$
to be consistent with the \rosat\ All-Sky Survey.
\end{abstract}

\section{Introduction}

The problem of the missing baryons is well known;
measurements of D in high redshift absorption line systems 
and application of Big Bang Nucleosynthesis models
reveal a baryon fraction, 
$\Omega h_{70}^2=0.0394\pm0.0029$, \citep{bt98}
while inventories of local material can account for only 
$\Omega h_{70}^{-1.5}=0.014\pm0.0034$ \citep{fhp98}.
Cosmological hydrodynamical simulations
(\citealt{ckor95}, \citealt{co99}, \citealt{bn98}, \citealt{dave2000}, 
and \citealt{cmdhkfw2000} for example)
suggest that the missing baryons
exist as shock-heated low-density filaments
of a ``warm-hot'' intergalactic medium
(WHIM) at temperatures of $\log T =5-7$.
The visibility of the WHIM is compromised by 
a number of Galactic foregrounds that exist
in the lower part of this temperature range,
the Local Hot Bubble ($\log T=6.11$, \citealt{sealhb98}) and
the Galactic Halo ($\log T=6.08$ and, perhaps $\log T=6.45$, \citealt{ks2000}).
All of these components emit predominately below E=1 keV,
where Galactic absorption can be significant.
A component at the high end of the WHIM range (e.g., $\log T=7$)
would emit strongly in the 1.5 keV band.
However, \citet{hea98} has directly resolved 70-80\%
of the flux in the 0.5-2.0 keV range,
leaving little room for diffuse components.
Thus, significant diffuse emission at the high end of the WHIM range
would be difficult to reconcile with observations,
while emission at the low end of the WHIM range becomes difficult to detect.

Without high resolution, high grasp, 
non-dispersive X-ray spectrometers,
there are three observational tests of the WHIM models;
the total emission, the energy spectrum,
and the fluctuation spectrum.
Theoretical calculations by \citet{co99} suggest that the WHIM produces 
$\sim7$ keV s$^{-1}$ cm$^{-2}$ sr$^{-1}$ keV$^{-1}$ at 0.7 keV;
the \citet{cmdhkfw2000} prediction for the entire IGM in the 0.5-2.0 keV band
yields the same value for the \tqkev\ band (\rosat\ R45 band).
\citet{poc2000}, however, estimate a substantially lower X-ray emission,
$\sim .7$ keV s$^{-1}$ cm$^{-2}$ sr$^{-1}$ keV$^{-1}$
in the 0.5-2.0 keV band.
\citet{dave2000} using a number of models including that of \citet{cmdhkfw2000}
predicts that the baryon fraction as a function of temperature
will be peaked near $\log T=6.45$.
Finally, \citet{cmdhkfw2000} predict that the value of the ACF
of the entire extragalactic SXRB in the 0.5-2.0 keV band
should be $\sim0.005$ for $\theta=10\arcmin$.

\section{The Emission Strength}

\begin{figure*}
\centerline{\psfig{figure=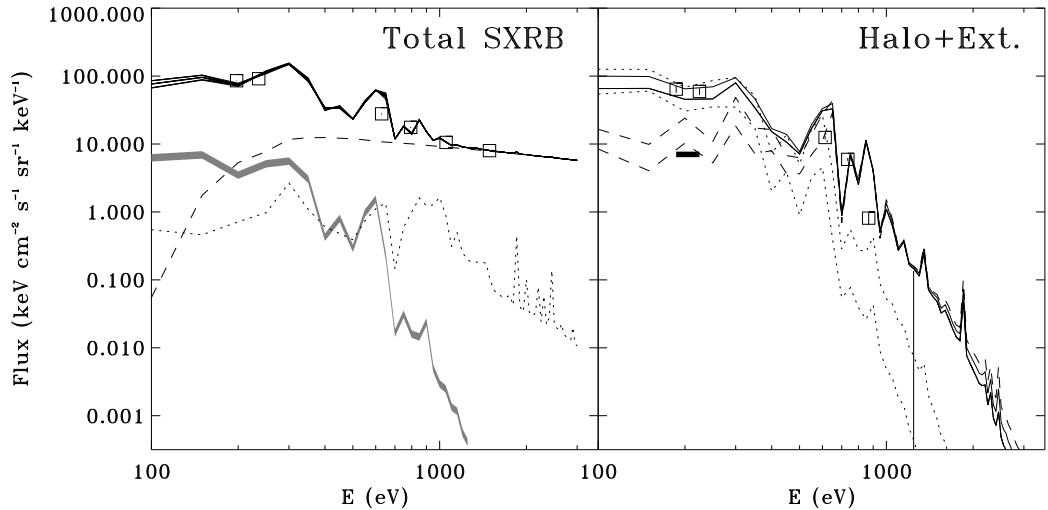,width=14.0cm}}
\caption{{\it Left:} The Total Soft X-ray Background.
The boxes are the \rass\ broad-band measurements.
The solid lines are our model with uncertainty at 100 eV resolution.
The dotted line is the contribution due to unresolved
Galactic stars after absorption by the Galactic disk.
The dashed line is the contribution by unresolved extragalactic sources
(the extragalctic power law) after absorption.
The shaded band is the contribution due to the LHB.
{\it Right:} The Galactic halo emission
and the extragalactic diffuse emission.
The dotted line is the spatially variable soft component
that is likely due to the halo.
The dashed line is the uniform hard component
which may contain contributions from both the Galactic halo
and extragalactic emission.
The bar is the maximum uniform \oqkev\ band emission
allowed by the measure of \citet{sfks2000}.
\label{fig:sed}}
\end{figure*}

To determine a limit to the strength of the WHIM emission,
we apply the methods of \citet{ks2000}
to the north Galactic polar cap,
$45\arcdeg<\ell<270\arcdeg$ and $b>85\arcdeg$.
The left panel of Figure~\ref{fig:sed} shows broad-band data
from the \rosat\ All-Sky Survey (\rass ),
for the north polar cap as well as the best fit model spectrum,
which is shown at a much higher energy resolution 
than measureable with \rosat .
The broad-band data has been corrected for
band-to-band point source detection limit differences.
We have assumed that the soft X-ray background (SXRB)
is composed of the following foregrounds/backgrounds
which may be successively removed.

{\it Unresolved Galactic Stars:}
The model of \citet{ks2001}
(similar to the model of \citet{ghmr96})
has been used to calculate the spectrum of the
Galactic stars below the \rass\ point source detection limit.
This model reproduces the number of stellar X-ray sources
in the $10^{-4}$ to $10^{-3}$ counts s$^{-1}$ range
found in deep pointings, to about 10\%;
unfortunately the luminosity function is expected to rise
for another two orders of magnitude,
so there may be larger uncertainties
for the total unresolved flux.

{\it Unresolved Extragalactic Point Sources:}
The measured spectrum of the unresolved extragalactic point sources
has been taken from \citet{cfg97} 
who find a photon index, $\Gamma=1.46$,
from individual \rosat\ and \asca\ spectra.
Since the \rass\ has a better non-cosmic background subtraction
than is possible for any single pointed observation,
the normalization of this extragalactic power-law
was adjusted so that all of the known emission components
account for all of the band R7 (1.05 keV $<E<$ 2.04 keV) flux.
The required normalization is 
$9.5\pm0.9$ keV s$^{-1}$ cm$^{-2}$ sr$^{-1}$ keV$^{-1}$,
the bulk of the uncertainty is due to the systematic uncertainty
in the R7 band flux, $\pm3.6\times10^{-6}$ counts s$^{-1}$ arcmin$^{-2}$.
This value is considerably less than the recent SAX value of
$11.7\pm0.5$ keV s$^{-1}$ cm$^{-2}$ sr${-1}$ keV$^{-1}$ \citep{vmgfp99},
and is $\sim18$\% higher than the HEAO-1 value of \citet{marshallea80}.
It has been assumed that the extragalactic power-law
does not have a break within the \rosat\ energy range
as \citet{mcba2000} find no break in the 0.3-10.0 keV range
for the aggregate spectrum of a sample
of their faintest {\it Chandra} sources.

{\it Local Hot Bubble:}
The strength of the LHB is determined from the variation
of observed flux with total absorbing column
in the manner of \citet{sealhb98} and \citet{ks2000}.

{\it The Remainder:}
The remaining flux contains contributions from all components
beyond the Galactic absorption 
that are not due to the unresolved extragalactic point sources;
the Galactic halo and any extragalactic diffuse emission.
The right panel of Figure~\ref{fig:sed} shows this remainder after correction
for the Galactic absorption.

As \cite{ks2000} showed that the emission from beyond the Galactic absorption
cannot be fit by a single thermal component,
we fit this remainder with two \citet{rs77} model components;
the lack of independent spectral resolution elements
prohibits fitting a greater number of components.
The fits to the north polar cap produce temperatures similar 
to those found in \citet{ks2000}
for $b>55\arcdeg,45\arcdeg<\ell<270\arcdeg$,
despite the differences in analysis
(subtraction of the composite stellar spectrum 
and the renormalization of the EPL);
$\log T_S=6.08_{-0.02}^{+0.04}$ and $\log T_H=6.43_{-0.01}^{+0.01}$.

The strength of the soft component varies strongly across the sky 
while the strength of the hard component is quite uniform \citep{ks2000},
at least at relatively high galactic latitudes ($|b|>30\arcdeg$).
The angular variation of the soft component suggests a Galactic origin,
while the uniformity of the hard component suggests 
that it is due either to
a Galactic corona in hydrostatic equilibrium, or extragalactic emission.
There is no way, on the basis of \rosat\ photometry,
to separate Galactic halo emission from diffuse extragalactic emission.
Therefore, the upper limit to the diffuse extragalactic emission
is the strength of the hard component,
$7.4\pm1.0$ keV s$^{-1}$ cm$^{-2}$ sr$^{-1}$ keV$^{-1}$ in the\tqkev\ band.
Note that the minimum \oqkev\ emission from beyond the Galactic absorption is
$\sim7$ keV s$^{-1}$ cm$^{-2}$ sr$^{-1}$ keV$^{-1}$ \citep{sfks2000},
consistent with the \oqkev\ flux produced by the hard component.

One might hope to place a limit on the contribution of the Galactic halo
by considering the halos of Milky Way analogues.
M101 is the closest face-on Milky Way analogue,
though its $D_{25}$ is about twice that of the Milky Way,
and its star formation rate as measured by $L_{FIR}/D_{25}^2$
is a factor of $\sim5$ lower than the Milky Way.
The sun is $\sim0.35D_{25}$ from the Galactic center;
at an equivalent radius in M101 
the upper limit of the \tqkev\ surface brightness due to M101
(which would include its halo and some emission from its disk)
is half the strength of our hard component in the same band.
However, for NGC 891, 
which has about half the mass of the Milky Way,
a similar $D_{25}$, and a similar $L_{FIR}/D_{25}^2$,
an observer at $\sim0.35D_{25}$ from its center would see
a galactic halo that is $\sim9$ times brighter 
than our hard component \citep{bh97}!
The relation between the strength of the halo
and the stellar formation rate is not sufficiently understood
to determine the fraction of the hard component that might be due
to our own Galactic halo.

\section{Energy Spectrum}

The fraction of baryons at each temperature in the \citet{cmdhkfw2000} model
is shown in \citet{dave2000}.
The baryon fraction at $z=0$ is peaked at $\log T\sim6.45$;
if the baryon fraction were weighted by the emission at each temperature,
the distribution would be more sharply peaked
and at somewhat higher temperatures.
The contribution of baryons at higher $z$
is substantially smaller than that of baryons at $z=0$
and is peaked at somewhat lower temperatures
($\log T\sim6.05$ at $z=1$).
Given the caveats above,
the agreement of the temperature of our hard component ($\log T=6.43$)
with the ``characteristic'' temperature predicted by the simulations
is comforting, but may be completely fortuitous
if the bulk of the emission is due to the Galactic halo.

\section{Fluctuation Spectrum}

We have determined the autocorrelation function (ACF)
for the \tqkev\ band (\rosat\ band R45)
from seven mosaics of deep PSPC pointings.
The non-cosmic background of each PSPC pointing
was removed using ESAS \citep{skman98},
all $4\sigma$ point sources were removed,
the image was restricted to the inner $53\arcmin$,
and the resulting data were mosaicked \citep{ksman98}.
The remaining non-cosmic background was removed
by comparing the mosaics to the \rass\ .
Using the variation of transmitted flux with column density,
we determined the strength of the LHB in the R1L2 band,
extrapolated this quantity to the R45 band,
and removed it from the image.
As the point source detection limit varies across each mosaic,
the mosaics were corrected using 
the extragalactic point source $\log N-\log S$ of \citet{hea98}
and the Galactic stellar luminosity function of \citet{ks2001}.
Each R45 band mosaic was then corrected for Galactic absorption.
The resultant image contains contributions from
the Galactic halo,
the unresolved extragalactic point sources,
and the extragalactic diffuse emission.
Once all the corrections were determined,
the mosaics were reconstructed using only the
central $32\arcmin$ radius region of each pointing.
The smallest scale for which one may meaningfully calculate the ACF
is the size of the largest PSF, $5\arcmin$.
(Although the \rosat\ PSF is $\sim0.25\arcmin$ on axis,
the size of the PSF grows rapidly with distance from the optical axis.)

For each mosaic we calculated the ACF,
\begin{equation}
W(\theta) = \frac{\sum\limits_{i,i^{\prime}} (R - \langle R \rangle)
(R^{\prime} - \langle R^{\prime} \rangle)
\sqrt{w w^{\prime}}}
{\left[\left({\sum\limits_i R + \sum\limits_{i^{\prime}} R^{\prime}}
\right)/2\right]^2 
\sum\limits_{i,i^{\prime}} \sqrt{w w^{\prime}}}
\end{equation}
where the sum is over all pairs of pixels seperated by $\theta$,
$R$ is the count rate in those pixels, and
$w$ is the statistical weight (the exposure time)
given to those pixel

\begin{figure*}
\centerline{\psfig{figure=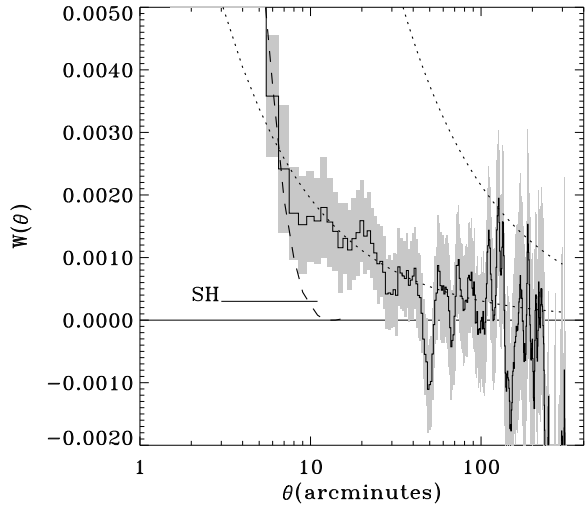,width=8.0cm}}
\caption{The ACF for the \tqkev\ band.
The ACF has been smoothed by a $3\arcmin$ sliding box.
The shaded region is the uncertainty in the ACF.
The dashed line is the maximum possible contribution to the ACF
by residual point sources.
The dotted lines are the ACF values of \citet{cfwbb94}
corrected from the R47 band to the R45 band.
The upper line is their measured ACF;
the lower line is their fitted ACF.
The horizontal line marked ``SH'' is the \citet{sh94} ACF value for the 1.5 keV 
band (\rosat\ band R67).
\label{fig:acf}}
\end{figure*}

We find that there is a significant mosaic-to-mosaic variation in the ACF,
which was confirmed by variance analyses ($\delta I/I$) of the mosaics.
The ACFs of individual mosaics were weighted by the mean exposure
to calculate the mean ACF shown in Figure~\ref{fig:acf}.
As the effective point source detection limit for a mosaic
can be much smaller than for individual exposures, 
the ACF contains a contribution from residual point sources
whose PSF radii range up to $5\arcmin$;
the maximum possible contribution due to the residual point sources
is shown by the dashed line.
Thus, the ACF shown is the upper limit for scales smaller than $10\arcmin$.

The short line labeled ``SH'' is the ACF determined by \citet{sh94}
for the \rosat\ R67 band (0.73 keV $<E<$ 2.04 keV);
over the interval shown it is nearly consistent with zero.
The dotted lines show the ACFs from \citet{cfwbb94},
corrected from their energy band (\rosat\ band R45 + R67)
to our band energy (\rosat\ band R45),
assuming that the fluctuations in the R67 band are negligible
for these scales.
The upper line is their measured ACF;
the lower line is the ACF that they derived by fitting
the data with the function $\alpha(\theta/1\arcmin)^{-0.8}$.

We chose to measure the ACF in the \tqkev\ band
as the hard component discussed in the previous section
will have a stronger contribution in this band at higher energies,
but there will not be as great a problem with Galactic absorption
as there is at lower energies.
However, the only published model ACF \citep{cmdhkfw2000}
is for the 0.5-2.0 keV band.
Since the \citet{sh94} ACF for higher energies
is much smaller than that for the \tqkev\ band,
the ACF for the 0.5-2.0 keV band can be estimated from
our ACF by scaling it by the quantity
\begin{equation}
\left[ \frac{\mbox{flux in our band}}
{\mbox{flux in 0.5-2.0 keV band}}\right] ^2,
\end{equation}
which is $\sim0.4$.
Taking the ACF for the $5\arcmin$ PSF mosaics,
we find that the ACF in the 0.5-2.0 keV band should be $\lesssim0.001$
for $10\arcmin < \theta < 20\arcmin$.
This is about a factor of five smaller than the model predictions.

There may, in fact, be contributions to the ACF
due to the Galactic halo and the LHB,
but these are expected to be small.
The ACF for emission from the LHB has been measured
for the $10\arcmin < \theta < 20\arcmin$ range
and is consistent with zero \citep{kuntz2000}.

Such a direct comparison of theory and observation is not correct.
The linear size of the \citet{cmdhkfw2000} simulation ($\sim1\arcdeg$)
corresponds to $\sim8$ Mpc$h^{-1}$ at a distance of 500 Mpc$h^{-1}$
and contains an X-ray emitting cluster of galaxies.
Since the mean separation of clusters is several times larger
than 8 Mpc$h^{-1}$, 
the simulation has an atypical overdensity 
(hence atypically large concentration of X-ray emission),
and thus a higher value of the ACF.

The observed ACFs show variation among mosaics with solid angles of
10 to 30 square degrees;
the root of the variance of our set of seven mosaics is about twice
the displayed uncertainty.
This variation may be consistent with cosmic variation.

\section{Discussion}

We have determined the upper limit to the strength
of diffuse extragalactic emission in the soft X-ray range,
by ``peeling off'' various known foregrounds and backgrounds.
What is left is some combination of extragalactic diffuse emission
and Galactic halo emission.
We expect our galaxy to have an X-ray halo
simply due to hydrostatic equilibrium
(the virial temperature is $\sim10^6$ K)
and because other galaxies like our own have them.
The strength of the halo remains a vexed issue.
The combined halo/extragalactic emission weighs in 
at just about the strength expected for the extragalactic component alone,
so either our galaxy has very little halo,
or the models over-predict the X-ray emission of the WHIM.
The lower \citet{poc2000} value suggests
that more care may be needed 
in comparing simulations to measurements.

It should be noted that this analysis has used a normalization
for the contribution by the unresolved extragalactic point souces
that is at the {\it lower} end of the range of values
commonly used.
Independent of the existence of extragalactic diffuse emission,
or the existence of a Galactic halo,
the \rass\ will not admit to a higher value for the normalization.
Increasing the normalization decreases the amount of flux
available for the extragalactic diffuse emission
to significantly smaller than that predicted by the models.

The upper limit to the ACF in the \tqkev\ band is smaller,
by about a factor of five,
than that predicted by the models at a scale of $10\arcmin$,
and all the improvements to the measurement of the ACF
(better correction for residual point sources and
more robust estimates of the contribution to the fluctuations
by the halo and the LHB)
will only further decrease the value of the ACF.
If a portion of the hard component strength is due to the galactic halo,
the weakness of the observed ACF signal
may be due to the reduced strength of the extragalactic emission.
However, as noted above, one should compare our mean ACF
with the ACF of a simulation whose size is comparable to
the cluster correlation length
rather than a small region surrounding a cluster.

\section{Conclusion}

The observed emission strength, energy spectrum, and fluctuation spectrum
of the soft X-ray background originating
from beyond the Galactic absorption
is roughly consistent with the predictions by
hydrodynamical simulations of the universe.
Even given the caveats concerning the Galactic halo,
the current observations do not {\it disprove}
the existing models.
The observations here present solid guideposts
for further exploration of the accessible parameter space
by future generations of simulations.

We wish to thank J.P. Ostriker and the anonymous referee
for their very useful comments.


\begin{thebibliography}{}

\bibitem[\protect\citeauthoryear{Bregman \& Houk}{Bregman \&
Houk}{1997}]{bh97}
Bregman, J.~N.,  \& Houk, J.~C. 1997, ApJ, 485, 159

\bibitem[\protect\citeauthoryear{Bryan \& Norman}{Bryan \&
Norman}{1998}]{bn98}
Bryan, G.~L.,  \& Norman, M.~L. 1998, ApJ, 495, 80

\bibitem[\protect\citeauthoryear{Burles \& Tytler}{Burles \&
  Tytler}{1998}]{bt98}
Burles, S.,  \& Tytler, D. 1998, ApJ, 499, 699

\bibitem[\protect\citeauthoryear{Cen et~al.}{Cen et~al.}{1995}]{ckor95}
Cen, R., Kang, H., Ostriker, J.~P.,  \& Ryu, D. 1995, ApJ, 451, 436

\bibitem[\protect\citeauthoryear{Cen \& Ostriker}{Cen \&
Ostriker}{1999}]{co99}
Cen, R.,  \& Ostriker, J.~P. 1999, ApJ, 514, 1

\bibitem[\protect\citeauthoryear{Chen, Fabian, \& Gendreau}{Chen
  et~al.}{1997}]{cfg97}
Chen, L.-W., Fabian, A.~C.,  \& Gendreau, K.~C. 1997, MNRAS, 285, 449

\bibitem[\protect\citeauthoryear{Chen et~al.}{Chen
et~al.}{1994}]{cfwbb94}
Chen, L.-W., Fabian, A.~C., Warwick, R.~S., Branduardi-Raymont, G.,  \&
Barber,
  C.~R. 1994, MNRAS, 266, 846

\bibitem[\protect\citeauthoryear{Croft et~al.}{Croft
  et~al.}{2000}]{cmdhkfw2000}
Croft, R. A.~C., Matteo, T.~D., Dav\'{e}, R., Hernquist, L., Katz, N.,
Fardel,
  M.~A.,  \& Weinberg, D.~H. 2000, ApJ, submitted

\bibitem[\protect\citeauthoryear{Dav\'e et~al.}{Dav\'e
et~al.}{2000}]{dave2000}
Dav\'e, R., et~al. 2000, ApJ, submitted

\bibitem[\protect\citeauthoryear{Fukugita, Hogan, \& Peebles}{Fukugita
  et~al.}{1998}]{fhp98}
Fukugita, M., Hogan, C.~J.,  \& Peebles, P. J.~E. 1998, ApJ, 503, 518

\bibitem[\protect\citeauthoryear{Guillout et~al.}{Guillout
  et~al.}{1996}]{ghmr96}
Guillout, P., Haywood, M., Motch, C.,  \& Robin, A.~C. 1996, A\&A, 316,
89

\bibitem[\protect\citeauthoryear{Hasinger et~al.}{Hasinger
  et~al.}{1998}]{hea98}
Hasinger, G., Burg, R., Giacconi, R., Schmidt, M., Tr\"umper, J.,  \&
Zamorani,
  G. 1998, A\&A, 329, 482

\bibitem[\protect\citeauthoryear{Kuntz}{Kuntz}{2000}]{kuntz2000}
Kuntz, K.~D. 2000, Ph.D. thesis, University of Maryland

\bibitem[\protect\citeauthoryear{Kuntz \& Snowden}{Kuntz \&
  Snowden}{1998}]{ksman98}
Kuntz, K.~D.,  \& Snowden, S.~L. 1998, Cookbook for analysis procedures
for
  {\it ROSAT} XRT Observations of Extended Objects and the Diffuse
Background,
  Part II: Mosaics, Technical report, NASA/GSFC

\bibitem[\protect\citeauthoryear{Kuntz \& Snowden}{Kuntz \&
  Snowden}{2000a}]{ks2000}
Kuntz, K.~D.,  \& Snowden, S.~L. 2000a, ApJ, 543, 195 

\bibitem[\protect\citeauthoryear{Kuntz \& Snowden}{Kuntz \&
  Snowden}{2000b}]{ks2001}
Kuntz, K.~D.,  \& Snowden, S.~L. 2000b, ApJ, in preparation

\bibitem[\protect\citeauthoryear{Marshall et~al.}{Marshall
  et~al.}{1980}]{marshallea80}
Marshall, F.~E., Boldt, E.~A., Holt, S.~S., Miller, R.~B., Mushotzky,
R.~F.,
  Rose, L.~A., Rothschild, R.~E.,  \& Serlemitsos, P.~J. 1980, ApJ, 235,
4

\bibitem[\protect\citeauthoryear{Mushotzky et~al.}{Mushotzky
  et~al.}{2000}]{mcba2000}
Mushotzky, R.~F., Cowie, L.~L., Barger, A.~J.,  \& Arnaud, K.~A. 2000,
Nature,
  404, 459

\bibitem[\protect\citeauthoryear{Phillips, Ostriker, \& Cen}{Phillips
  et~al.}{2000}]{poc2000}
Phillips, L.~A., Ostriker, J.~P.,  \& Cen, R. 2000, ApJ, submitted

\bibitem[\protect\citeauthoryear{Raymond \& Smith}{Raymond \&
  Smith}{1977}]{rs77}
Raymond, J.~C.,  \& Smith, B.~W. 1977, ApJS, 35

\bibitem[\protect\citeauthoryear{Snowden et~al.}{Snowden
  et~al.}{1998}]{sealhb98}
Snowden, S.~L., Egger, R., Finkbeiner, D., Freyberg, M.~J.,  \&
Plucinsky,
  P.~P. 1998, ApJ, 493, 715

\bibitem[\protect\citeauthoryear{Snowden et~al.}{Snowden
  et~al.}{2000}]{sfks2000}
Snowden, S.~L., Freyberg, M.~J., Kuntz, K.~D.,  \& Sanders, W.~T. 2000,
ApJS, 128, 171


\bibitem[\protect\citeauthoryear{Snowden \& Kuntz}{Snowden \&
  Kuntz}{1998}]{skman98}
Snowden, S.~L.,  \& Kuntz, K.~D. 1998, Cookbook for analysis procdures
for {\it
  ROSAT} XRT Observations of Extended Objects and the Diffuse
Background, Part
  I: Individual observations, Technical report, NASA/GSFC

\bibitem[\protect\citeauthoryear{Soltan \& Hasinger}{Soltan \&
  Hasinger}{1994}]{sh94}
Soltan, A.,  \& Hasinger, G. 1994, A\&A, 288, 77

\bibitem[\protect\citeauthoryear{Vecchi et~al.}{Vecchi
et~al.}{1999}]{vmgfp99}
Vecchi, A., Molendi, S., Guainazzi, M., Fiore, F.,  \& Parmar, A.~N.
1999,
  A\&A, 349, L73

\end{thebibliography}

\end{document}